\documentclass[twocolumn,floatfix,superscriptaddress,showpacs,showkeys,nofootinbib]{revtex4}
\usepackage{epsfig}
\usepackage{amssymb,latexsym,amsmath}
\newcommand{\eq}[1]{\begin{align} #1 \end{align}}

\begin{document}

\title{Bose-Einstein Condensation of Pions in High Multiplicity Events}

 \author{V.V. Begun}
 \affiliation{Museo Storico della Fisica e Centro Studi e Ricerche
 Enrico Fermi, Rome, Italy}
 \affiliation{Bogolyubov Institute for Theoretical Physics, Kiev, Ukraine}
 \author{M.I. Gorenstein}
 \affiliation{Bogolyubov Institute for Theoretical Physics, Kiev, Ukraine}
 \affiliation{Frankfurt Institute for Advanced Studies, Frankfurt,Germany}

\begin{abstract}
We present microcanonical ensemble calculations of particle number
fluctuations in the ideal pion gas approaching Bose-Einstein
condensation. In the samples of events with a fixed number of all
pions, $N_{\pi}$, one may observe a prominent signal. When
$N_{\pi}$ increases the scaled variances for particle number
fluctuations of both neutral and charged pions increase
dramatically in the vicinity of the Bose-Einstein condensation
line. As an example, the estimates are presented for $p+p$
collisions at the beam energy of 70~GeV.

\end{abstract}



\maketitle

The phenomenon of Bose-Einstein condensation (BEC) was predicted
long time ago \cite{Bose}. Tremendous efforts were required to
confirm BEC experimentally. The atomic gases are transformed into
a liquid or solid before reaching the BEC point. The only way to
avoid this is to consider extremely low densities. At these
conditions the thermal equilibrium in the atomic gas is reached
much faster than the chemical equilibrium. The life time of the
{\it metastable} gas phase is stretched to seconds or minutes.
This is enough to observe the BEC signatures. Small density leads,
however, to small temperature of BEC. Only in 1995 two
experimental groups succeeded to create the `genuine' BE
condensate  by using new developments in cooling and trapping
techniques \cite{cold_atoms}. Leaders of these two groups,
Cornell, Wieman, and Ketterle, won the 2001 Nobel Prize for this
achievement.

Pions are spin-zero mesons. They are the lightest hadrons
copiously produced in high energy collisions. In the present
letter we argue that the pion number fluctuations may give a
prominent signal of approaching the BEC point. In fact, there is
the BEC line in a plane of pion density and temperature. The pion
system should be in a {\it metastable} state (in thermal, but not
chemical equilibrium) to reach the BEC line. This can be achieved
by selecting the samples of events with high pion multiplicities.
Multipion states are formed in high energy nucleus-nucleus
collisions, as well as in the elementary particle ones. There were
several suggestions to search for BEC of $\pi$-mesons (see, e.g.,
Ref.~\cite{pion-BC}). However, complete statistical mechanics
calculations of pion number fluctuations have never been
presented. There is a qualitative difference in properties of the
mean multiplicity and of the scaled variance of multiplicity
fluctuations in different statistical ensembles. The results
obtained with grand canonical ensemble (GCE), canonical ensemble
(CE), and microcanonical ensemble (MCE) for the mean multiplicity
approach to each other in the large volume limit. This reflects
the thermodynamic equivalence of the statistical ensembles.
Recently it has been found  \cite{CE,BF,MCE} that corresponding
results for the scaled variance are different in different
ensembles, and this difference is preserved in the thermodynamic
limit. To extract the matter properties from analysis of
event-by-event fluctuations, one needs to fix the samples of high
energy events, and choose the corresponding statistical ensemble
for their analysis. This is discussed below.

Let us start with a well known example of non-relativistic ideal
Bose gas. The occupation numbers, $n_{\bf{p}}$, of single quantum
states, labelled by 3-momenta $\bf{p}$, are equal to
$n_{\bf{p}}=0,1,\ldots,\infty$. In the GCE their average values,
fluctuations, and correlations are the following  \cite{lan}:
\eq{
 &\langle n_{\bf{p}} \rangle
  = \frac {1} {\exp \left[\left(\frac{{\bf p}^2}{2m}~ -~ \mu \right)/ T\right]
 ~-~ 1}~,~~
 \nonumber
 \\
 &\langle\left(\Delta n_{\bf{p}}\right)^2\rangle
 = \langle n_{\bf{p}}\rangle \left(1
+ \langle n_{\bf{p}} \rangle\right)\equiv
\upsilon^{2}_{\bf{p}}~,~~
\nonumber
 \\
 & \langle \Delta n_{\bf{p}}  \Delta n_{\bf{k}} \rangle
 =\upsilon_{\bf{p}}^2~\delta_{\bf{p}\bf{k}}~, %
\label{mcc-gce}
} where $\Delta n_{\bf{p}}\equiv n_{\bf{p}}-\langle
n_{\bf{p}}\rangle$, $m$ denotes the particle mass, $T$ and $\mu$
are the system temperature and chemical potential, respectively
(throughout the paper we use the units with $\hbar=c=k=1$).
The average number of particles in the GCE reads \cite{lan}:
\eq{
\langle N\rangle~ &\equiv~ \overline{N}(V,T,\mu)~=~\sum_{\bf{p}}
\langle n_{\bf{p}} \rangle\nonumber
 \\
 &=~\frac{V}{2\pi^2}\int_0^{\infty}\frac{p^2dp}
 {\exp\left[\left(\frac{p^2}{2m}~-~\mu \right)/T\right]~-~1}~,
\label{N-nonrel}
}
where $V$ is the system volume. We consider particles with spin
equal to zero, thus the degeneracy factor equals 1.  In the
thermodynamic limit, $V\rightarrow \infty$, the sum over momentum
states is transformed into the momentum integral,
$\sum_{\bf{p}}\ldots =(V/2\pi^2)\int_0^{\infty}\ldots p^2dp$. This
substitution, assumed in all formulae below,  is valid  if the
chemical potential in the non-relativistic Bose gas is restricted
as $\mu < 0$ (or $\mu <m$ in relativistic formulation). When the
temperature $T$ decreases at fixed ratio, $\langle N\rangle/V$,
the chemical potential $\mu$ increases and becomes equal to zero
at $T=T_{C}$, known as the BEC temperature. At this point from
Eq.~(\ref{N-nonrel}) one finds,
$\overline{N}(V,T\!\!=\!\!T_{C},\mu\!\!=\!\!0)\!=\!V[mT_{C}/(2\pi)]^{3/2}\zeta(3/2)$,
where $\zeta(3/2)\cong 2.612$ is the  Riemann zeta-function. At
$\mu=0$ and $T<T_{C}$, a macroscopic part, $N_0$ (called the BE
condensate), of the total particle number occupies the lowest
energy level $p=0$.

Introducing $\Delta N \equiv N-\langle N\rangle$ one finds the
particle number fluctuations in the GCE,
\eq{ \label{dNdN}
& \langle (\Delta N)^2 ~\rangle ~=~\sum_{\bf{p},\bf{k}}~\langle
\Delta n_{\bf{p}}~\Delta n_{\bf{k}}\rangle~=~\sum_{\bf{p}}~
v_{\bf{p}}^2~, \\
%
%
%
%
%
& \omega \equiv \frac{\langle (\Delta N)^2\, \rangle}{\langle N
\,\rangle} \,=\, \frac{\sum_{\bf{p}}
\upsilon_{\bf{p}}^2}{\sum_{\bf{p}}\langle n_{\bf{p}}\rangle}
~=~1~+~\frac{\sum_{\bf{p}}\langle
n_{\bf{p}}\rangle^2}{\sum_{\bf{p}}\langle n_{\bf{p}}\rangle}~.
\label{omega-nonrel}
}
The limit $-\mu/T\gg 1$ gives $\langle n_{\bf{p}}\rangle\ll 1$.
This corresponds to the Boltzmann approximation, and then from
Eqs.~(\ref{N-nonrel},\ref{omega-nonrel}) it follows:
$\overline{N}(V,T,\mu)\cong V\exp(\mu/T)(mT/2\pi)^{3/2}$ and
$\omega\cong 1$. When $\mu$ increases the scaled variance $\omega$
becomes larger, $\omega>1$. This is the well known Bose
enhancement effect for the particle number fluctuations. From
Eq.~(\ref{omega-nonrel}) at $\mu\rightarrow 0$ one finds
$\omega\rightarrow\infty$. Two comments are appropriate. First,
for finite systems $\omega$ remains finite, and $\omega = \infty$
emerges from Eq.~(\ref{omega-nonrel}) at $\mu=0$  in the
thermodynamic limit $V\rightarrow\infty$, when the sums over $\bf
{p}$ are transformed into the momentum integrals. Second, the
anomalous fluctuations of the particle number at the BEC point
correspond to the GCE description. In the CE and MCE, the number
of particles $N$ in a non-relativistic system is fixed by
definition, thus, $\omega_{c.e.}=\omega_{m.c.e.}=0$. At $\mu=0$
and $T<T_C$ the average number of particles in the BE condensate
is proportional to $1-(T/T_C)^{3/2}$ and it remains the same in
the CE and MCE at $V\rightarrow \infty$. The fluctuations of $N_0$
are, however, very different and strongly suppressed in the CE and
MCE \cite{BE-fluc}.

We consider now the relativistic ideal gas of pions,
%
\eq{
\langle n_{{\bf p},j}\rangle~=~\frac{1}{\exp[(\sqrt{{\bf p^2}+m_{\pi}^2}~-~\mu_j)/T]~-~1}~,
\label{np-pions}
}
where index $j$ enumerates 3 isospin pion states, $\pi^+,\pi^-$,
and $\pi^0$, the energy of one-particle states is taken as,
$\epsilon_{\bf p}=({\bf p^2}+m_{\pi}^2)^{1/2}$ with $m_{\pi}\cong
140$ MeV being the pion mass (we neglect a small difference
between the masses of charged and neutral pions).
The inequality $\mu_j\le m_{\pi}$ is a general restriction in the
relativistic Bose gas, and $\mu_{j}=m_{\pi}$ corresponds to the
BEC.
In Ref.~\cite{BF} we discussed in details the Bose gas with one
conserved charge in the CE $(V,T,Q=const)$, i.e. the
$\pi^+\pi^-$-gas with fixed electric charge. This corresponds to
the GCE $(V,T,\mu_Q)$, thus, in Eq.~(\ref{np-pions}) $\mu_+
=\mu_Q$ and $\mu_-=-\mu_Q$ for $\pi^+$ and $\pi^-$, respectively.
Approaching the BEC of $\pi^+$ at $\mu_Q\rightarrow m_{\pi}$, one
finds the relation between $T_C$ and $\rho_Q\equiv\rho_+-\rho_-$
(the picture of BEC of $\pi^-$ at $Q<0$ and $\mu_Q\rightarrow
-m_{\pi}$ is obtained by a mirror reflection). At $T_C/m_{\pi}\ll
1$ it coincides with the non-relativistic formula, $T_C\cong 3.31
\,\rho_Q^{2/3}m_{\pi}^{-1}$, and at $T_C/m_{\pi}\gg 1$ it gives
$T_C\cong 1.73\,\rho_Q^{1/2}m_{\pi}^{-1/2}$ (see, e.g.,
Ref.~\cite{kapusta}). The scaled variance $\omega^+\equiv\langle
(\Delta N_+)^2\rangle/\langle N_+\rangle$ in the GCE goes to
infinity. This is similar to the non-relativistic case. On the
other hand, the scaled variance for negative particles,
$\omega^-\equiv\langle (\Delta N_-)^2\rangle/\langle N_-\rangle$,
remains finite and even decreases with $\mu_Q$. The pion numbers
$N_+$ and $N_-$ fluctuate in the both GCE and CE. However, the
exact conservation imposed in the CE on the system charge,
$Q=N_+-N_-$, suppresses anomalous fluctuations at the BEC point:
$\omega_{c.e.}^+$ is finite with the upper limit,
$\zeta(2)/\zeta(3)\cong 1.368$, reached at $T_{C}/m\rightarrow
\infty$ (see details in Ref.~\cite{BF}).

In what follows we discuss a rather different pion system.
In the MCE $(V,E,Q=0,N_{\pi}=const)$ formulation, the total system
energy $E$, electric charge $Q=N_+ -N_-$, and total number of
pions $N_{\pi}\!=\!N_0\!+\!N_+\!+\!N_-$ will be fixed. This
corresponds to the GCE~$(V,T,\mu_Q\!\!=\!0,\mu_{\pi})$ description
with $\mu_+=\mu_{\pi}+\mu_Q$, $\mu_-=\mu_{\pi}-\mu_Q$, and
$\mu_0=\mu_{\pi}$ in Eq.~(\ref{np-pions}). We restrict
$\mu_Q\!\!=\!0$ and consider BEC when
$\mu_{\pi}\!\rightarrow\!m_{\pi}$. The $\mu_Q\!\!=\!0$ corresponds
to zero electric charge, $Q\!\!=\!0$ or $N_+=N_-$, in the pion
system.

\begin{figure}[t!]
 \hspace{-0.45cm}
 \epsfig{file=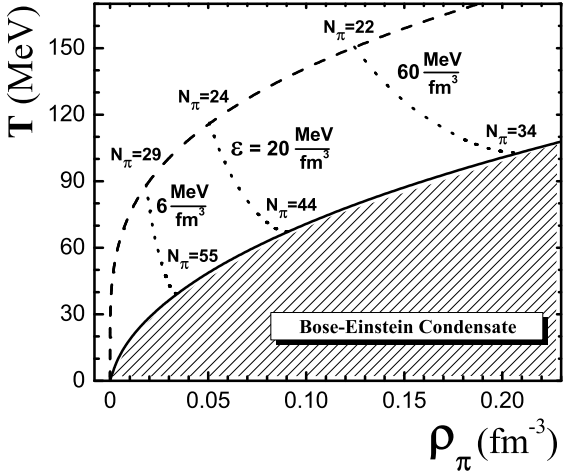,width=0.47\textwidth}
 \vspace{-0.6cm}
 \caption{
 The phase diagram of the pion gas with $\mu_Q=0$. The dashed line
 corresponds to $\rho_{\pi}(T,\mu_{\pi}\!\!=0)$, and  the solid line
 to BEC (\ref{T_BC}). The dotted lines show the states
 with fixed energy densities:
 $\varepsilon = 6, 20, 60$ MeV/fm$^3$.
 The $N_{\pi}$  numbers in the figure correspond to $\mu_{\pi}=0$
 and $\mu_{\pi}=m_{\pi}$ at these energy densities for the total pion energy,
 $E~=~9.7$~GeV.
 \label{fig-cond}}
\vspace{-0.5cm}
\end{figure}

The pion density is equal to
$\rho_{\pi}(T\!,\mu_{\pi})\!=\!\sum_{{\bf p},j} \langle n_{{\bf
p},j}\rangle/V$. The phase diagram of the ideal pion gas in
$\rho_{\pi}-T$ plane is presented in Fig.~\ref{fig-cond}. BEC
starts at  $T=T_{C}$ when $\mu_{\pi}=\mu^{max}_{\pi}=m_{\pi}\;$.
It gives:
 \eq{\label{T_BC}
  &\rho_{\pi}(T=T_C,\mu_{\pi}=m_{\pi})
  \\
 &= \frac{3\,T_{C}\,m_{\pi}^2}{2\pi^2}
 \sum_{n=1}^{\infty}\frac{1}{n}\,
 K_2\left(n\,m_{\pi}/T_{C}\right)\,\exp(n\,m_{\pi}/T_{C}) ,\nonumber
  }
where $K_2$ is the modified Hankel function. The Eq.~(\ref{T_BC})
gives the BEC line shown by the solid line in Fig.~\ref{fig-cond}.
If $T_{C}/m_{\pi}\ll 1$, from Eq.~(\ref{T_BC}) one finds,
$T_{C}\cong
2\pi\,[3\zeta(3/2)]^{-2/3}\,\rho_{\pi}^{2/3}m_{\pi}^{-1}$. This
corresponds to the non-relativistic limit discussed above, but
with a degeneracy  factor $g_{\pi}\!=\!3$. In the
ultrarelativistic limit, $T_{C}/m_{\pi}\!\gg\!1$, from
Eq.~(\ref{T_BC}) it follows:
$T_{C}\!\cong\![\pi^2/3\zeta(3)]^{1/3}\rho_{\pi}^{1/3}$. We
consider the region in $\rho_{\pi}-T$ plane between the
$\mu_{\pi}\!=\!0$ and $\mu_{\pi}\!=\!m_{\pi}$ lines. The lines of
fixed energy density, $\varepsilon(T,\mu_{\pi})=\sum_{{\bf
p},j}\epsilon_{{\bf p}}\,\langle n_{{\bf p},j}\rangle/V$, are
shown as dotted lines in Fig.~\ref{fig-cond} inside this region
for three fixed values of $\varepsilon$. An increase of
$\rho_{\pi}$ at constant $\varepsilon$ leads to the increase of
$\mu_{\pi}$ and decrease of $T$. In this letter we discuss how the
system approaches the BEC line
$(\mu_{\pi}\!=\!m_{\pi},T\!=\!T_{C})$, and do not touch the region
$(\mu_{\pi}\!=\!m_{\pi},T\!<\!T_{C})$ below this line where the
non-zero BE condensate is formed. The GCE~$(V,T,\mu_Q,\mu_{\pi})$,
MCE~$(V,E,Q,N_{\pi})$, and CE~$(V,T,Q,N_{\pi})$ are equivalent for
average quantities, including average particle multiplicities, in
the thermodynamic limit. Thus, Eq.~(\ref{T_BC}) and phase diagram
in Fig.~1 remain the same in all statistical ensembles. However,
the pion number fluctuations are very different in different
ensembles.

As an example we consider the high multiplicity events in $p+p$
collisions at IHEP (Protvino) accelerator with the beam energy of
70~GeV (see Ref.~\cite{Dubna} on the experimental project
``Thermalization'', team leader V.A.~Nikitin). In  the reaction
$p+p\rightarrow p+p+N_{\pi}$ with small final proton momenta in
the c.m.s.,  the total c.m. energy of created pions is $E\cong
\sqrt{s}-2m_{p}\cong 9.7$~GeV. The trigger system designed at JINR
(Dubna) selects the events with $N_{\pi}>20$ in this reaction.
This makes it possible to accumulate the samples of events with
fixed $N_{\pi}=30\div 50$ and the full pion identification during
the next 2 years \cite{nikitin}. Note that for this reaction the
kinematic limit is $N_{\pi}^{max}=E/m_{\pi}\cong 70$. The pion
system in the thermal equilibrium is expected to be formed for
high multiplicities. The volume of the pion gas system is
estimated as, $V=E/\varepsilon(T,\mu_{\pi})$, and the number of
pions equals to $N_{\pi}= V\rho_{\pi}(T,\mu_{\pi})$. The values of
$N_{\pi}$ at $\mu_{\pi}=0$ and $\mu_{\pi}=m_{\pi}$ for 3 different
values of energy density $\varepsilon$ are shown in
Fig.~\ref{fig-cond}.

For $Q=0$, the average pion multiplicities, $\langle N_0 \rangle
=\langle N_{\pm} \rangle =N_{\pi}/3$, are the same in all
statistical ensembles for large systems. This thermodynamic
equivalence is not, however, valid for the scaled variances of
pion fluctuations. The system with the fixed electric charge,
$Q=0$, the total pion number, $N_{\pi}$, and total energy of the
pion system, $E$, should be treated in the MCE. The volume $V$ is
one more (and unknown) MCE  parameter. The calculations below are
carried out in a large volume limit, thus, parameter $V$ does not
enter explicitly in the formulae for the scaled variances. The
microscopic correlators for the MCE ($V, E, Q=0, N_{\pi}$) equal
to:
%
 \eq{ \label{mcorr}
& \langle \Delta  n_{{\bf p},j} \Delta n_{{\bf k},i}  \rangle_{m.c.e.}
  =  \upsilon_{{\bf p},j}^2~\delta_{{\bf p}{\bf k}}\delta_{ji}
  -  \upsilon_{{\bf p},j}^2\,\upsilon_{{\bf k},i}^2 \\
 ~&\times \left[ \frac{q_jq_i}{\Delta (q^2)}\nonumber
  +   \frac{\Delta (\epsilon^2)
   +  \epsilon_{\bf{p}}\epsilon_{\bf{k}}\; \Delta (\pi^2)
   -  (\epsilon_{\bf{p}} + \epsilon_{\bf{k}})\Delta (\pi\epsilon)}
 {\Delta (\pi^2)\Delta (\epsilon^2)  -  (\Delta (\pi\epsilon))^2}
 \right]\, ,
  }
where $q_+\!\!=\!1,\,q_-\!\!=\!-1,\,q_0\!\!=\!0$; $\upsilon_{{\bf
p},j}^2\!\!=\!(1+\langle n_{{\bf p},j}\rangle)\langle n_{{\bf
p},j}\rangle$; $\Delta(q^2)\!\!=\!\!\sum_{{\bf p},j}q_j^2
\upsilon_{{\bf p},j}^2,\;
 \Delta (\pi^2)\!\!=\!\!\sum_{{\bf p},j} \upsilon_{{\bf p},j}^2,\;
\Delta (\epsilon^2)\!\!=\!\!\sum_{{\bf p},j} \epsilon_{\bf{p}}^2
\upsilon_{{\bf p},j}^2,\;
 \Delta (\pi \epsilon)\!\!=\!\!
   \sum_{{\bf p},j} \epsilon_{\bf{p}} \upsilon_{{\bf p},j}^2$.
Note that the first term in the r.h.s. of Eq.~(\ref{mcorr})
corresponds to the GCE. Correlations between differently charged
pions, $j\!\neq\! i$,  and between different single modes, ${\bf
p}\!\ne\!{\bf k}$, are absent in the GCE. It then follows:
%
%
$\omega^+ = \omega^- = \omega^0
 \equiv \omega = 1 +  \sum_{\bf{p}}\langle
n_{\bf{p}}\rangle^2/\sum_{\bf{p}}\langle n_{\bf{p}}\rangle~,
%
%
%
$ similar to  Eq.~(\ref{omega-nonrel}), but with $\langle
n_{\bf{p}}\rangle = \{\exp[(\sqrt{{\bf
p^2}+m_{\pi}^2}-\mu_{\pi})/T]~-~1\}^{-1}$. In the GCE the numbers
$N_+$, $N_-$, and $N_0$ fluctuate independently of each other. The
Bose effects in the pion system are small if $\mu_{\pi}=0$. One
finds $\omega=1.01\div 1.12$ in the temperature interval $T=40\div
160$~MeV (note that $\omega=1$ in the Boltzmann approximation).
The Bose effects increase with $\mu_{\pi}$, and $\omega\rightarrow
\infty$ at $\mu_{\pi}\rightarrow m_{\pi}$, i.e. approaching the
BEC line the GCE calculations give anomalous fluctuations for
$N_+$, $N_-$, and $N_0$. The MCE $(V,E,Q=0,N_{\pi}=const)$
formulation means the restrictions of the fixed total system
energy $E$, electric charge $Q=N_+ -N_-$, and total number of
pions $N_{\pi}=N_0+N_+ +N_-$ for the each microscopic state of the
system.
\begin{figure}[t!]
\vspace{-0.2cm}
 \hspace{-0.45cm}
 \epsfig{file=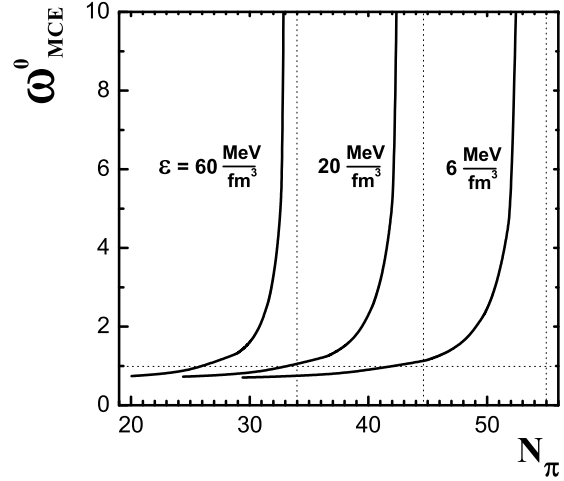,width=0.47\textwidth}
 \vspace{-0.6cm}
  \caption{
The scaled variance of neutral pions in the MCE  is presented as
the function of the total number of pions. Three solid lines
correspond to different energy densities: $\varepsilon = 6, 20,
60$ MeV/fm$^3$. The total energy of the pion system is assumed to
be fixed, $E= 9.7$~GeV. The vertical dotted lines correspond to
the points on the BEC line at the specific values of the energy
density.
 \label{fig-w}
 }
\vspace{-0.5cm}
\end{figure}
This changes the pion number fluctuations. From Eq.~(\ref{mcorr})
one notices that the MCE fluctuations of each mode ${\bf p}$ are
reduced, and the (anti)correlations between different modes ${\bf
p}\ne {\bf k}$ and between different charge states appear. This
results in a suppression  of all scaled variances
$\omega_{m.c.e.}$ in comparison with the corresponding ones
$\omega$ in the GCE. Note that the MCE microscopic correlators
(\ref{mcorr}), although being different from that in the GCE, are
expressed with the quantities calculated in the GCE. The MCE
scaled variances depend on two GCE parameters: $T$ and
$\mu_{\pi}$.  The straightforward calculations lead to the
following MCE scaled variance:
\eq{
 \omega^0_{m.c.e.} \;=\;
 \frac{\sum_{\bf{p},\bf{k}}\langle
 \Delta n_{\bf{p}}^0~\Delta n_{\bf{k}}^0
 \rangle_{m.c.e.}}{\sum_{\bf{p}}\langle n_{\bf{p}}^0\rangle}~=~
 \frac{2}{3}~\omega~.
 \label{omega-mce}
 }
Due to conditions, $N_+\equiv N_-$ and $N_++N_-+N_0\equiv
N_{\pi}$, it follows,
$\omega_{m.c.e.}^{\pm}\!\!\!=\!\omega_{m.c.e.}^0/4$ and
$\omega_{m.c.e.}^{ch}\!\!\!=\!\omega_{m.c.e.}^0/2$, where
$N_{ch}\equiv N_++N_-$. The behavior of $\omega^0_{m.c.e.}$ is
shown in Fig.~\ref{fig-w}. To make a correspondence with $N_{\pi}$
values, we consider again the $p+p \rightarrow p+p+N_{\pi}$
collisions at the beam energy of 70~GeV and take  the pion system
energy to be equal to $E\!=\!9.7$ GeV. Despite of the MCE
suppression the scaled variances for the number fluctuations of
$\pi^0$ and $\pi^{\pm}$ increase dramatically and abruptly when
the system approaches the BEC line.

As an instructive example  let us consider the MCE
$(V,E,Q\!\!=\!0,N_{ch}=const)$, i.e. fixed $N_{ch}$ instead of
$N_{\pi}$. The corresponding GCE formulation gives the following
pion chemical potential: $\mu_+=\mu_{\pi}$, $\mu_0=0$, $\mu_-=
\mu_{\pi}$ in Eq.~(\ref{np-pions}) ($\mu_Q=0$, as before, because
of $Q=0$ condition). When $\mu_{\pi}\rightarrow m_{\pi}$ the
system approaches the BEC line for $\pi^+$ and $\pi^-$. The
thermodynamic behavior and position of this BEC line can be easily
found. Approaching the BEC line one can also find
$\omega^{\pm}_{g.c.e.}\rightarrow\infty$. The pion number
fluctuations are, however, very different in the  MCE
$(V,E,Q\!\!=\!0,N_{ch})$.
In the statistical ensembles with fixed $N_{ch}$ and $Q$ no
anomalous BEC fluctuations are possible. The numbers of $N_+$ and
$N_-$ are completely fixed by the conditions $Q=N_+-N_-=0$ and
$N_{ch}=N_++N_-=const$, thus,
$\omega_{c.e.}^{\pm}=\omega_{m.c.e.}^{\pm}=0$. The number $N_0$
fluctuates, but $\mu_0=0$, thus, neutral pions are far away from
the BEC line and their fluctuations are small, $\omega^0\approx 1$
in all statistical ensemble formulations.

The broad distributions over $N_0$ and $N_{ch}$ close to the BEC
line also implies large fluctuations of the $f\equiv N_0/N_{ch}$
ratio. These large fluctuations were suggested (see, e.g.,
Ref.~\cite{DCC}) as a possible signal for the disoriented chiral
condensate (DCC). The DCC leads to the distribution of $f$ in the
form, $dW(f)/df=1/(2\sqrt{f})$. The thermal Bose gas corresponds
to the $f$-distribution centered at $f=1/3$. Therefore,
$f$-distributions from BEC and DCC are very different, and this
gives a possibility to distinguish between these two phenomena.

The calculations presented in this letter should be improved by
taking into account the finite size effects, pion-pion
interactions, and some other effects. However, the described BEC
scenario may survive the complications.  A crucial point is the
analysis of the samples of high $N_{\pi}$ events.  The following
inequalities are always hold for particle number fluctuations in
different ensembles:
$\omega^i_{m.c.e.}<\omega^i_{c.e.}<\omega^i_{g.c.e.}$. Therefore,
if the anomalous BEC fluctuations are present in the MCE, they are
also exist (and even larger) in the CE and GCE. The reverse
statement is not true. The anomalous BEC fluctuations of the GCE
may disappear in the CE or MCE. We found that for the system with
$N_{\pi}=const$ and $Q=0$  the anomalous BEC fluctuations do not
wash out by exact conservation laws of the CE and MCE. The
required $N_{\pi}$ values for the BEC are much larger than the
average pion multiplicity per collision, thus, these high
$N_{\pi}$ events are rather rare and give negligible contributions
to inclusive observables in high energy collisions.
With increasing of $N_{\pi}$ in the sample with fixed total
energy, the temperature of the pion system has to decrease and it
approaches the BEC line. This can happen in different ways: at
constant energy density $\varepsilon$, at constant pion density
$\rho_{\pi}$, or with decreasing of both $\varepsilon$ and
$\rho_{\pi}$. The pion system should move to the BEC line one way
or another. In the vicinity of the BEC line (no BE condensate is
yet formed) one observes an abrupt and anomalous increase of the
scaled variances of neutral and charged pion number fluctuations.
This could (may be even should) be checked experimentally.

{\bf Acknowledgments.} We would like to thank F.
Becattini, K.A.~Bugaev, A.I.~Bugrij, I.M.~Dremin, M.~Ga\'zdzicki,
W.~Greiner, K.A.~Gridnev, M.~Hauer, I.N.~Mishustin,
St.~Mr\'owczy\'nski and Yu.M.~Sinyukov for discussions and
comments. We are also grateful to E.S.~Kokoulina and V.A.~Nikitin.
They informed us on the experimental project \cite{Dubna}, and
this stimulated the present study. We thank S.V.~Chubakova for
help in the preparation of the manuscript. The work was supported
in part by US CRDF, 
project agreement UKP1-2613-KV-04, and by Ukraine-Hungary
cooperative project M/101-2005.

\end{document}